# Objective Assessment of Corneal Transparency in the Clinical Setting with Standard SD-OCT Devices

**Maëlle Vilbert,**[1,4] **Romain Bocheux,**[1,4,5] **Cristina Georgeon,**[4] **Vincent Borderie,**[4] **Pascal Pernot,**[2] **Kristina Irsch,**[3,4‡] **Karsten Plamann**[1,5‡]

[1] Laboratory for Optics and Biosciences (LOB) — École polytechnique, CNRS, INSERM, IPP, Palaiseau, France
[2] Physical Chemistry Institute (ICP) — CNRS, University of Paris-Saclay, Orsay, France
[3] Vision Institute — CNRS, INSERM, Sorbonne University, Paris, France
[4] GRC 32, Transplantation et Thérapies Innovantes de la Cornée, Sorbonne Université, Centre Hospitalier National d'Ophtalmologie des Quinze-Vingts
[5] LOA — ENSTA Paris, École polytechnique, CNRS, IPP, Palaiseau, France
‡ Joint Senior Authors

Correspondence: Maëlle Vilbert, Laboratoire d'Optique et Biosciences (U1182, UMR7645) – École polytechnique, Route de Saclay, 91128 Palaiseau Cedex, FRANCE; maelle.vilbert@polytechnique.edu.

## ABSTRACT

**PURPOSE**: To develop an automated algorithm allowing extraction of quantitative corneal transparency parameters from clinical spectral-domain OCT images. To establish a representative dataset of normative transparency values from healthy corneas.

**METHODS**: SD-OCT images of 83 normal corneas (ages 22–50 years) from a standard clinical device (RTVue-XR Avanti, Optovue Inc.) were processed. A pre-processing procedure is applied first, including a derivative approach and a PCA-based correction mask, to eliminate common central artifacts (i.e., apex-centered column saturation artifact and posterior stromal artifact) and enable standardized analysis. The mean intensity stromal-depth profile is then extracted over a 6-mm-wide corneal area and analyzed according to our previously developed method deriving quantitative transparency parameters related to the physics of light propagation in tissues, notably tissular heterogeneity (Birge ratio; $B_r$), followed by the photon mean-free path ($l_s$) in homogeneous tissues (i.e., $B_r \sim 1$).

**RESULTS**: After confirming stromal homogeneity ($B_r < 10$, IDR: 1.9–5.1), we measured a median $l_s$ of 570 μm (IDR: 270–2400 μm). Considering corneal thicknesses, this may be translated into a median fraction of transmitted (coherent) light $T_{coh(stroma)}$ of 51% (IDR: 22–83%). No statistically significant correlation between transparency and age or thickness was found.

**CONCLUSIONS**: Our algorithm provides robust and quantitative measurement of corneal transparency from standard clinical SD-OCT images. It yields lower transparency values than previously reported, which may be attributed to our method being exclusively sensitive to spatially coherent light. Excluding images with central artifacts wider than 300 μm also raises our median $T_{coh(stroma)}$ to 70% (IDR: 34–87%).

# INTRODUCTION

The clinical evaluation of corneal transparency usually consists of a qualitative and undetailed inspection of opacities using a slit-lamp biomicroscope, sometimes with comparison against a subjective and arbitrary grading scale (e.g., 0 to 4 or 5) [1,2]. While several approaches have been proposed to quantify and/or objectively assess corneal transparency [3] (e.g., via slit-lamp biomicroscopy [3,4], the Scheimpflug principle [5], confocal microscopy [6], and optical coherence tomography (OCT) [7]), none has been found suitable to gain widespread usage. Hence, there remains a clinical need for a reliable and easy-to-use objective approach that enables quantitative corneal transparency characterization, including monitoring ability, towards effective prevention, diagnosis, and treatment of various pathologies.

Our team previously developed an objective method for quantitative corneal transparency assessment based on a new optical data analysis-based approach, which was validated *in vitro* using high-resolution three-dimensional data obtained by full-field optical coherence tomography (FF-OCT) [8]. Briefly, our approach consists of:

(1) Calculating an in-depth average of the OCT signal;
(2) Estimating its signal-to-noise-ratio (SNR) by comparing it to a low-frequency spline interpolation;
(3) Calculating a mono-exponential function fitted to the stromal data;
(4) Calculating the Birge ratio representing the model adequacy.
    a. For corneas characterized as homogeneous by low Birge ratios, we extract numerical values for the photon mean-free path length and its confidence interval;
    b. For corneas characterized as heterogeneous by elevated Birge ratios, we calculate the depth-dependent variations about the mean-free path length.

The present study translates this work into clinical practice by applying our method to standard spectral-domain (SD-)OCT devices providing two-dimensional (B-scan) images. The specific challenges of this clinical application include the much wider field-of-view (FOV) of these devices, and the variations due to the positioning of the patient with respect to the FOV and the fixed focal plane within a B-scan image (in contrast to FF-OCT, where the relatively small FOV images are always acquired in the focal plane). Our approach thus pays particular attention to the detection and compensation of common acquisition artifacts and includes an automated pre-processing algorithm that computes a correction mask.

# METHODS

*Instrumentation and image acquisition*
2D OCT (B-scan) images from a standard clinical SD-OCT device (RTVue-XR Avanti OCT; Optovue Inc., Fremont, CA, USA) were used in this study. The SD-OCT device included the cornea-anterior module lens (CAM), specifically the wide-angle lens (CAM-L) configuration (allowing for essentially telecentric scanning across the cornea). Acquisition rates were 26,000 A-scans/second, with 256 to 1024 A-scans/frame. The SLD was centered at $840 \pm 10$ nm with a spectral bandwidth $\Delta\lambda = 50$ nm (FWHM); light exposure at the pupil was $750$ µW. The axial and transverse (theoretical) resolutions in tissue were $5$ µm and $15$ µm, respectively. The lateral FOV was 6 to 9 mm, depending on the acquisition mode, and the axial scanning range was 2 to 2.3 mm. The axial image pixel size was calibrated using the automated central pachymetry (= central corneal thickness, CCT) measurements provided by the Optovue software (v. 2018.1.1.63). The raw images were stored as 8-bit grayscale PNG files and anonymized after manual USB export.

*Patient selection for creation of a normative database*

SD-OCT images of $n = 83$ normal corneas from 43 subjects aged $31 \pm 13$ years were included in this study, which was approved by the Institutional Review Board (Patient Protection Committee, Île-de-France V) and adhered to the tenets of the Declaration of Helsinki. The criterion for study inclusion was registration for, but before, refractive surgery at the Quinze-Vingts National Eye Hospital's anterior segment department in the time frame from 2018 to 2021. Images were collected retrospectively; however, all patients provided informed oral consent to have their images used in research. Two raw B-scans per cornea, corresponding to horizontal (nasal-temporal) section views, were processed (i.e., a total of 166 B-scans), one from cross-sectional views ('Line' or 'CrossLine' mode) and the other from pachymetry maps ('Pachymetry' or 'PachymetryWide' mode).

For comparison and discussion purposes, we also analyzed SD-OCT images from 2 pathological corneas with compromised transparency as per "gold-standard" subjective and qualitative image inspection following the same institutional and ethics guidelines.

Similarly, to get an idea of the repeatability of our method, 2 normal corneas (from the same adult subject) were measured prospectively, with informed consent, 40 times by the same observer on the same day, 10 times per OCT acquisition mode (namely 'Cross', 'Line', 'Pachy', 'PachyWide'). Intraclass correlation coefficient estimates ($ICC_{3,k}$ and $ICC_{3,1}$) and their 95% confidence intervals [8] were calculated using the Python programming language (Python Software Foundation, v2.7.4) and 'irr' package [9] based on a two-way mixed-effects model, consistency and multiple/single measurements for each OCT acquisition mode — the acquisition modes being considered as fixed raters in the model.

Note that for the presentation of our results, 95% confidence intervals ($CI_{95}$) of normally distributed data are determined according to Student's t-distribution coefficient $t$; for a given variable $x$, $CI_{95} = \bar{x} \pm t\, SD$ where SD denotes the standard deviation. For samples with size $n > 30$, $t = 1.96$. For non-normally distributed data, the interdecile range (IDR; the range between the 10th percentile and the 90th percentile) is provided.

*Pre-processing algorithm*

A pre-processing algorithm was developed in Python (v2.7.4) and will be soon available at the following open-source repository: https://github.com/maelle-v. This procedure is performed to standardize the raw OCT images and to extract the in-depth attenuation profile from the corneal stroma while compensating for acquisition artifacts.

We differentiate between two "hyperreflective" artifacts, observed centrally, which are associated with the instrumental configuration and patient positioning and are often termed central artifacts [10]. (1) A prominent central artifact, affecting the entire apex-centered column of the image (that we term saturation artifact; Figure 1A and 1B), with strong back-reflection of the incident light at the air-tear(-epithelial) interface (as well as the endothelial-aqueous interface), which saturates the line camera of the SD-OCT's spectrometer and results in a hyporeflective region between the corneal surfaces. The artifact's periodic pattern (not observed with swept-source OCT systems) is due to the harmonics generated by the Fourier transform (employed as part of the SD-OCT signal reconstruction) of the sharp-edged shape of the saturated interferograms [11]. (2) A less prominent central artifact (Figure 1C), with a hyperreflective region in the posterior stroma (that we term posterior stromal artifact; yellow arrow in Figure 1C) that interestingly is not observed in SD-OCT images of a corneal phantom containing spherical shaped scatterers (Cornea model eye, Rowe Technical Design Inc., Dana Point, CA, USA; Figure 1D) and hence may only result from signal reconstruction in a layered medium (e.g., stromal lamellae in real corneas), and be slightly reinforced by corneal refraction.

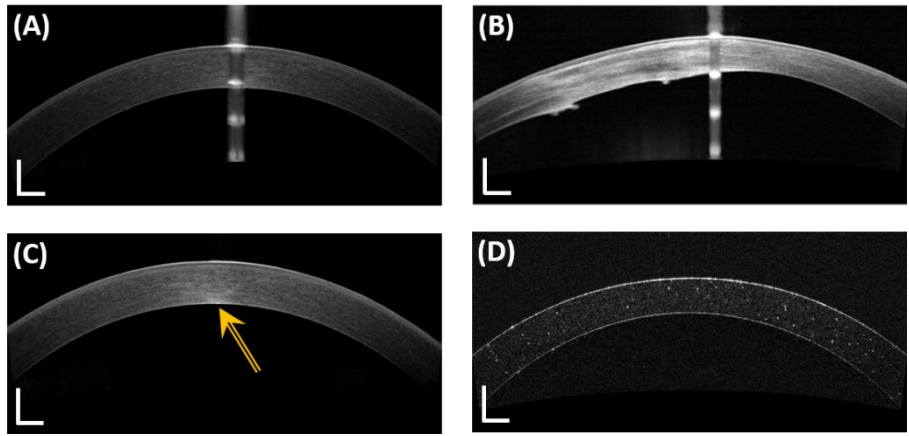

*Figure 1. Central artifacts observed in clinical SD-OCT images of corneas* (RTVue-XR Avanti OCT; Optovue Inc., Fremont, CA, USA). *(A-B, Saturation artifact) Prominent central artifact with a repetitive pattern of back-reflections from the air-tear(-epithelial) interface and endothelium-aqueous interface in the example of a (A) normal cornea and (B) pathological cornea. (C, Posterior stromal artifact) Less prominent central artifact with a hyperreflective region in the posterior stroma (yellow arrow), not observed in (D) SD-OCT image of a corneal phantom (Cornea model eye, Rowe Technical Design Inc., Dana Point, CA, USA). Scale bar lengths: 500 µm.*

The automated steps of the pre-processing algorithm are illustrated in Figure 2. Each step, associated with a sub-figure, is detailed below:

A. **Raw image input and optional histogram sliding**. If the imported raw image is overexposed (mean histogram value is higher than a given threshold), a custom consistent value is subtracted from each pixel of the image to **restore a standard exposure**.

B. **Detection and optional removal of the saturation artifact** (see Figure 1A, 1B). The key feature here is the relatively high intensity of the few neighboring columns in the artifact zone. The signal level of each column (A-scan) is averaged to get the mean A-scan intensity as a function of the lateral (x-) coordinate. This signal is smoothened ($2^{nd}$ order Savitzky-Golay filter, 15-pixel window to remove discontinuities) and its first derivative is computed. If the absolute value of this derivative reveals an abrupt change in lateral (x-) intensity, the presence of a prominent central artifact is confirmed. In that case, the lateral boundaries of the artifact are set at the x-coordinates corresponding to the two extreme values of the derivative (maximum derivative value for the left boundary, minimum value for the right) ± a 55-µm margin. The OCT image is then cropped according to those boundaries. User consent is required; if denied, a cursor enables manual (x-) segmentation of this artifact.

C. **Detection of anterior surface**. For each column of the smoothened image (Gaussian filter, $\sigma = 2$), the local maximum corresponding to the depth (z-) coordinate of the air-tear film interface is determined based on an SNR-dependent thresholding procedure. In case of detection error at the edges of the image, the anterior surface is linearly interpolated and subsequently smoothened (median filter, 15-pixel window).

D. **Flattening of the cornea and stromal segmentation**. Each column of the exposure-adjusted image is axially translated so that all anterior surface z-coordinates (located in step C) horizontally match. Each line of the flattened cornea is then averaged, to get the mean in-depth OCT signal. The three main peaks of this signal are detected, which correspond to the z-coordinates of the following interfaces: air-tear, epithelium-epithelial basement membrane/Bowman's layer, and endothelium-aqueous. The **region of interest (ROI) delineates stromal boundaries**; it is determined for an apex-centered 6-mm-wide lateral area (x-axis) and lies between the epithelial basement membrane (EBM) and the endothelium (z-axis). The ROI excludes two axial (z-) margins: a 50-µm margin after the epithelium-EBM/Bowman's layer interface as well as a 30-µm margin before the endothelium-aqueous interface, to assure exclusion of both Bowman's and Descemet's layers contribution, respectively. The (z-) margins were chosen empirically based on values known from corneal histology that were further adjusted to optimize the repeatability of our transparency measures for normal corneas (see below). User consent is required; if denied, a cursor enables manual segmentation of the stromal ROI.

E. **Lateral localization of the posterior stromal artifact** (yellow arrow in Figure 1C). The segmented stromal ROI is numerically divided into 20 sub-layers of a constant thickness (see Figure 2E, top) and color-coded according to depth (i.e., the warmer the color, the deeper the sub-layer). The mean lateral (x-) OCT intensity of each sub-layer is then computed, as illustrated in Figure 2E, bottom (using the same color-coding as in Figure 2E, top). The individual values are stored in a matrix and a principal component analysis (PCA) is performed (the associated eigenvectors can indeed depict data tendencies and discrepancies; see Supplementary Methods for more details), using Python's *sklearn.decomposition* library; the lateral (x-) coordinates of the artifact zone are located from the standard deviation of the $2^{nd}$ principal component eigenvalues (as indicated by the dashed red lines in Figure 2E).

F. **Computation and application of a customized correction mask** to compensate for the posterior stromal artifact. A second PCA is performed on the same input data in the artifact zone (i.e., the specifically located x-coordinate range); the intensity corresponding to a customized correction mask for the posterior stromal artifact is derived from the $1^{st}$ principal component eigenvalues of this second PCA (see Figure 2F, top). Upon application of the mask, one can observe in Figure 2F, bottom, that the intensities of the deepest sub-layers have been numerically attenuated (in comparison to Figure 2E, bottom).

G. **Normalization of the image** using the smoothed intensity signal at the anterior surface ($2^{nd}$ order Savitzky-Golay filtering, 115-pixel window). Every value below 75 on an 8-bit grayscale is set equal to 75, to avoid overemphasizing the contribution of edges. This step compensates for any lateral illumination variability related to the geometry of the corneal curvature.

H. **Averaging of the final image along the depth direction** (*z*-axis), to extract the mean intensity at a given stromal depth. This stromal in-depth attenuation profile is exported as a CSV file.

The mean computation time for the entire pre-processing algorithm is around 2.3 seconds, excluding the GUI interaction time dedicated to user approval (94-bit OS, Inter® Core™ i7-8665U, CPU at 1.9 GHz and 16 GB DDR4 RAM).

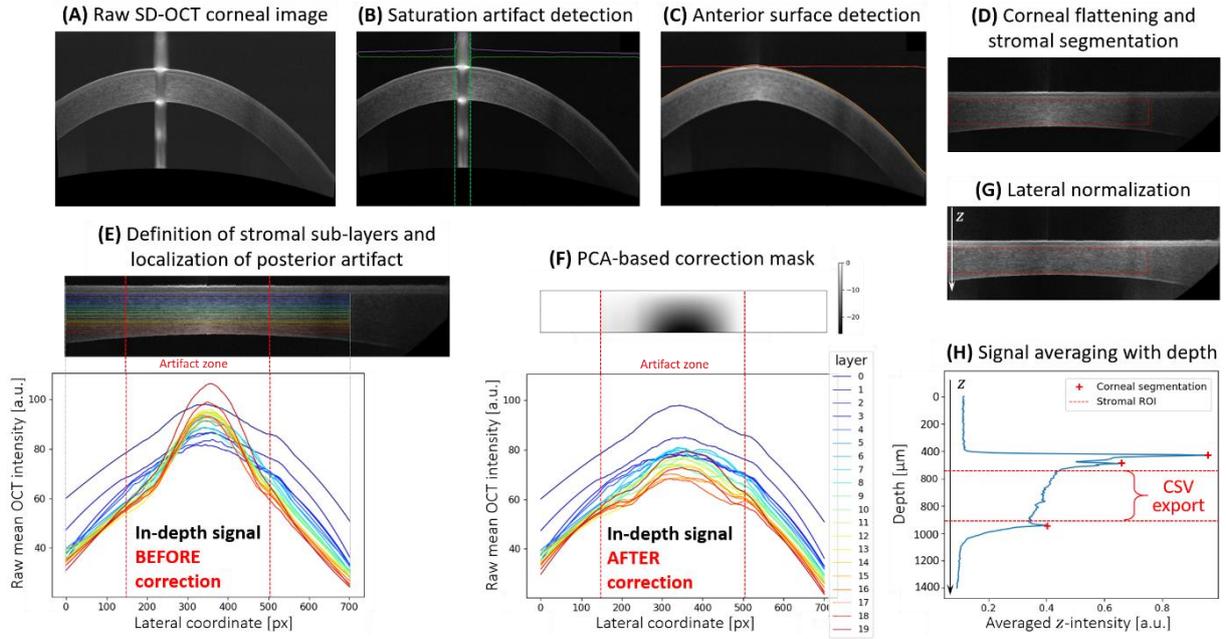

*Figure 2. Graphical representation of the pre-processing algorithm with individual steps for SD-OCT corneal image standardization and stromal in-depth intensity profile extraction. (A) Raw SD-OCT cross-sectional 'Line' corneal image. (B) Saturation artifact detection after histogram sliding to adjust exposure. (C) Detection of anterior surface. (D) Numerical flattening of the cornea and segmentation of the apex-centered stromal region of interest (ROI). (E, F) Computation of a customized correction mask to account for the posterior stromal artifact. (G) Lateral normalization of the flattened and corrected image. (H) Extraction of the stromal in-depth intensity profile.*

*Fitting algorithm and extraction of quantitative transparency parameters*

The averaged stromal backscattering profile is in turn analyzed using our developed Bayesian approach-based algorithm available here: https://doi.org/10.5281/zenodo.2579915 [12] (R Core Team, v3.6.3), which has been described in detail previously [8]. Briefly, a variety of objective parameters, relevant for light propagation in tissues and transparency assessment, are derived from fitting a mono-exponential decay model to the OCT signal as a function of stromal depth. Among the objective parameters are the **signal-to-noise-ratio (SNR)** of the pre-processed attenuation (stromal $z$-) profile, the **Birge ratio ($B_r$)**, equivalent to a reduced $\chi^2$, that quantifies stromal homogeneity (homogeneous if $B_r \sim 1$, inhomogeneous if $B_r \gg 1$), and the **scattering mean-free path ($l_s$**; a major indicator of scattering extent and thus of transparency of a medium).

Note that in the normal (homogeneous) stroma, the incoming coherent wavefront is exponentially attenuated by scattering processes (following a Lambert-Beer law). The propagation distance corresponding to an attenuation by a factor $1/e$ is called the scattering mean-free path ($l_s$). Given that the average OCT signal at any given depth is proportional to the intensity of the incident wavefront, its measurement permits the assessment of $l_s$.

$l_s$ together with the stromal thickness (more specifically, the stromal ROI) permits to calculate the **fraction of the transmitted coherent wavefront** as:

$$T_{coh(stroma)} = \exp\left(-\frac{\text{stromal thickness}}{l_s}\right).$$

## RESULTS

*Reliability assessment*

Profile SNR varies among the acquisition modes, given the negative lower boundary of their respective intraclass correlation coefficient (ICC) 95% confidence intervals (see Supplementary Table 1 for ICC measurement details [13]). The values obtained for the SNR are specific to the acquisition mode (see also the following sub-section for more detail).

Tissue homogeneity and transparency measures, namely $B_r$, $l_s$, and $T_{coh(stroma)}$, have excellent inter-mode reliability in 'Line' and 'Cross' OCT acquisition modes for the mean of $k$ measurements (p-values << 0.05), with respective ICC$_{3,k}$ [CI$_{95}$] of 0.999 [0.91—1], 0.999 [0.98—1] and 0.999 [0.96—1]. For pachymetry mapping OCT images ('Pachy' and 'PachyWide' modes), on the other hand, we found a fixed bias for $T_{coh(stroma)}$ measurements, according to pairwise comparisons with the Tukey Honest Significant Difference (HSD) test (see Supplementary Figure 1 for Bland-Altman comparison of OCT acquisition modes); that bias was significant in 'Pachy' mode, which can also be noted in Figure 3, where the measurements for the right eye (OD; right panel) are significantly higher in 'Pachy' mode than in the other modes.

Reliability is excellent for single measurements of $B_r$, $l_s$ and $T_{coh(stroma)}$ in 'Line' and 'Cross' OCT acquisition modes; the associated ICC$_{3,1}$ [CI$_{95}$] estimates are 0.999 [0.84—1], 0.995 [0.95—1] and 0.999 [0.92—1] for $B_r$, $l_s$ and $T_{coh(stroma)}$, respectively.

When performing multiple measurements in 'Line' or 'Cross' mode, we observed that the average sample coherent transmittance converges within a $\pm 5\%$ interval for a sample size of at least 3 similar images (same OCT acquisition mode) and within a $\pm 3\%$ interval for a sample size of 5 images or more (see illustration on Figure 3). The respective results for 'Pachy' and 'PachyWide' modes are 6 similar images or more for a $\pm 5\%$ interval and at least 8 images for a $\pm 3\%$ interval.

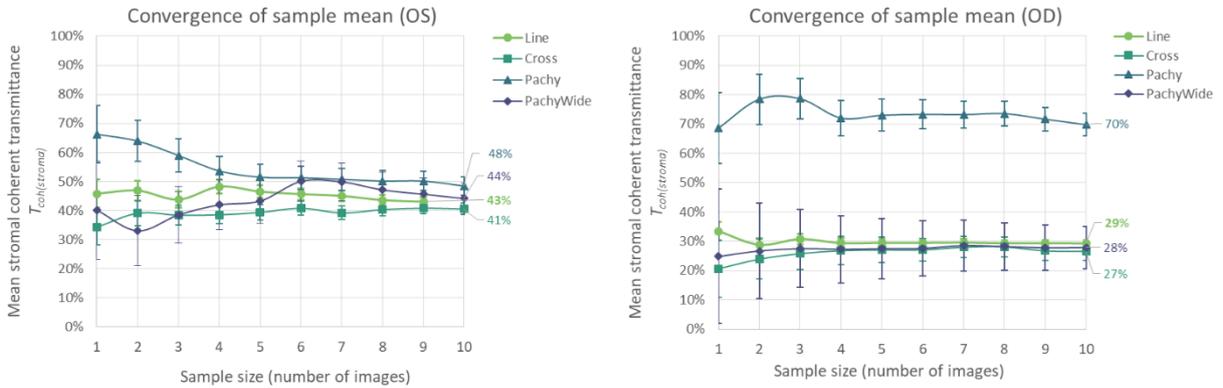

*Figure 3. Convergence of the sample mean $T_{coh(stroma)}$ according to the sample size. The sample size corresponds to the number of analyzed images from the same eye, acquired at the same moment by the same observer. The graphs depict the data for the left eye (OS; left panel) and right eye (OD; right panel) tested for reliability.*

Moreover, we found no significant bias from the removal of the saturation artifact in the raw images (Pearson's correlation test between artifact width and $T_{coh(stroma)}$, normally distributed: p-value >> 0.05 for every acquisition mode).

*Image-quality and stromal-homogeneity assessment*

The quality of raw images, characterized by a bidimensional signal-to-noise ratio (SNR$_{2D}$), depends on the acquisition mode (see Figure 4, left). There is a significant difference of ~3 dB between cross-sectional images ('Line', 'Cross') and pachymetry mapping images ('Pachy', 'PachyWide'). However, this difference in image quality was found to have no significant impact on the quality (SNR) of the in-depth stromal attenuation profile extracted from the raw images (mean difference < 0.4 dB; see Figure 4, middle); all extracted profiles can theoretically be analyzed with the fitting algorithm.

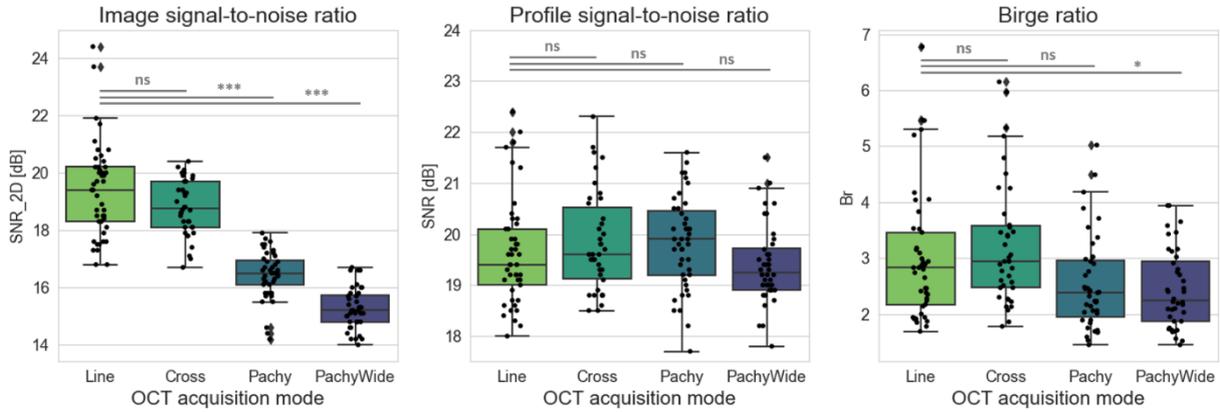

*Figure 4. Image quality and stromal homogeneity derived from SD-OCT images: $SNR_{2D}$ of raw images, SNR of 1D attenuation profiles, and Birge ratio ($B_r$) of the exponential fit.* The results of pairwise comparisons (with a Tukey HSD posthoc test) are represented above the box plots; "ns" stands for non-significant, while p-values (P) lower than 0.05 are considered as significant (* stands for $P < 0.05$, ** for $P < 0.01$, and *** for $P < 0.001$). A total of 166 B-scans, were analyzed, of which 45 were Line scans, 38 CrossLine scans (here 'Cross'), 43 Pachymetry scans (here 'Pachy'), and 40 PachymetryWide scans (here 'PachyWide').

Calculated Birge ratios vary slightly with the OCT acquisition mode, but remain rather close to unity ($B_r < 10$; IDR$_{all\ modes}$: 1.8–4.1, IDR$_{'Line'\&'Cross'}$: 1.9–5.1) compared to heterogeneous corneas ($B_r \gg 1$; [14,15]). Hence every normal cornea included in our study is considered to have a homogeneous stroma and can be used to establish a representative dataset of normative transparency values via the calculation of $l_s$.

*Determination of normative transparency values*

Even with every available OCT attenuation profile having a sufficient SNR and an acceptable Birge ratio, we established our normative database for corneal transparency measures using only high-quality and high-reliability images acquired in cross-sectional modes, that is, 'Line' ($n = 45$) and 'Cross' ($n = 38$) OCT scans. Results are shown in Figure 5, with details on the distributions in Supplementary Figure 2. The precision (Δ) of single measurements is derived from the dispersion of the data for reliability testing ($\Delta x = t$ SD with $t = 2.26$ being the coefficient given by the Student's t-distribution for $n = 10$ images and 95% confidence interval), namely $\Delta l_s = \pm 120$ μm and $\Delta T_{coh(stroma)} = \pm 9\%$ for the 'Line' mode and $\Delta l_s = \pm 230$ μm and $\Delta T_{coh(stroma)} = \pm 18\%$ for the 'Cross' mode.

We obtain a lognormal distribution of photon mean-free path values, with mean value $\exp[\overline{\log(l_s)}] = 710$ μm, median($l_s$) = 570 μm, min($l_s$) = 190 μm, max($l_s$) = 4100 μm, and interdecile range (IDR): $270 - 2400$ μm.

These translate into a bimodal distribution for the fraction of transmitted coherent light, with median($T_{coh(stroma)}$) = 51%, mean($T_{coh(stroma)}$) = 53%, min($T_{coh(stroma)}$) = 13%, max($T_{coh(stroma)}$) = 90%, IDR: 22–83%, and bimodal peaks around 38% and 80%. The corneal thickness in our group ranges from 469 μm to 602 μm (CI$_{95}$), and the associated mean percentage of analyzed stromal depth (in-depth stromal ROI) corresponds to 74% (SD = 4%) of the full corneal thickness.

We observe an impact of the removal of the saturation artifact on the measured coherent transmittance: the wider the saturated zone, the lower $T_{coh(stroma)}$ (Spearman's rank correlation test between artifact width and $T_{coh(stroma)}$: $\rho = -0.44$, p-value $= 10^{-5}$). This impact is significant for central cuts larger than 300 µm, as illustrated in Figure 6.

In our group of patients, we observe no significant correlation between coherent transmittance and age or corneal thickness (Spearman's rank correlation test: $\rho = 0.10$, p-value $= 0.39$ for age and $\rho = 0.17$, p-value $= 0.12$ for corneal thickness, respectively); similarly, no significant correlation between age and corneal thickness was observed (Spearman's rank correlation test: $\rho = -0.02$, p-value $= 0.85$). These conclusions remain the same when tests are performed on the reduced sample ($n = 42$) of images with saturation artifact cuts narrower than 300 µm.

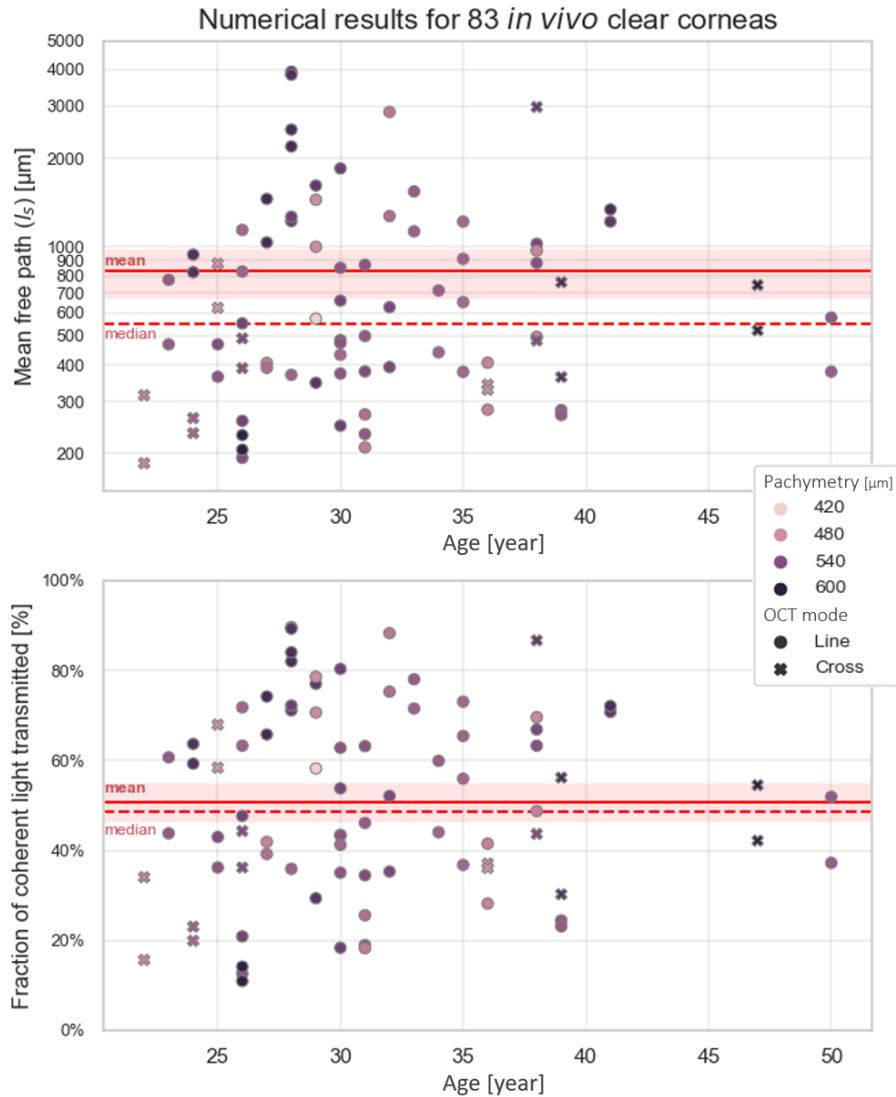

*Figure 5. Transparency measures of $n = 83$ normal corneas*, with subject age as x-axis and central pachymetry (CCT) as hue: (top) $l_s$, (bottom) $T_{coh(stroma)}$. Red shades correspond to 1.96 times the standard error (SE = SD/$\sqrt{n}$) on both sides of the mean. Precision is ±120 µm for $l_s$ measurements and ±9% for $T_{coh(stroma)}$ measurements in 'Line' mode, and respectively $\Delta l_s$ =±230 µm and $\Delta T_{coh(stroma)}$ = ±18% in 'Cross' mode. No correlation was found between transparency and age or corneal thickness in our group of patients.

Finally, we note that the grouping factor at the subject level (considering n = 80 corneas from 40 eye pairs) explains 98% [96–99%] of the total variance of central corneal thickness (CCT) and 55% [30–73%] of the total variance of $T_{coh(stroma)}$ measurements (ICC$_{1,1}$ computation based on a one-way random effects model, for absolute agreement and single measurement: p-values $\ll 0.05$). In other words, two normal corneas from a single person are 98% more likely to have equal thickness and 55% more likely to have similar transparency measures than if they are taken at random. Similarly, there is no significant difference between left and right eyes in CCT or transparency measurements, for images with saturation artifact narrower than 300 µm (Wilcoxon's signed-rank test: W = 13, p-value = 0.86 for CCT and W = 60, p-value = 0.66 for $T_{coh(stroma)}$).

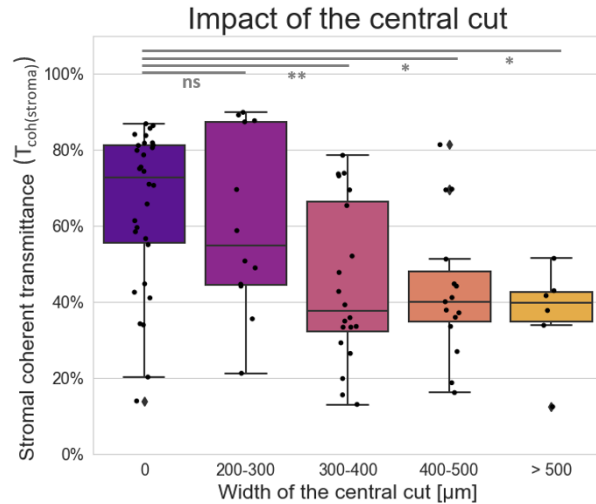

*Figure 6. Impact of saturation-artifact removal on the stromal coherent transmittance. The "central cut" corresponds to the part of the image that has been cropped after detection of a saturated area (step B in the pre-processing algorithm, see Figure 2). The central cut width is either null or superior to 200 µm. The results of pairwise comparisons with Tukey HSD posthoc test are represented above the box plots; "ns" stands for non-significant, while p-values (P) lower than 0.05 are considered as significant (\* stands for $P < 0.05$, and \*\* for $P < 0.01$).*

## DISCUSSION

The automated algorithm described in this article enables the extraction of quantitative parameters related to corneal transparency from clinical OCT images acquired with a standard SD-OCT device while eliminating common instrument-related artifacts.

Based on the analysis of typical corneal B-scans acquired in routine clinical practice, our method was found to have excellent reliability when applied to cross-sectional images (i.e., 'Line' and 'Cross' modes of the RTVue-XR Avanti OCT by Optovue Inc.). Other raw images, such as the ones used for corneal thickness mapping (e.g., 'Pachy', 'PachyWide') turned out to be a source of fixed biases: their poorer SNR (due to a lower number of averaged B-scans per final image) is associated with an uneven sampling of the in-depth signal notably in the anterior stroma, which in turn may lead to inaccurate fitting in that region. Consequently, pachymetry maps raw images should not be used in the context of the current method. The precision of our measurements on cross-sectional images (i.e., 'Line' and 'CrossLine') could be statistically improved by analyzing several same-eye OCT images and averaging the associated results, as illustrated in Figure 3 (resulting in dividing the uncertainty of the mean value by the square root of the number of scans), as has been previously done to improve the repeatability of quantitative measurements with the RTVue OCT system by Optovue [16]. For example, by averaging the measures derived from 3 'Line' or 'Cross' OCT images, we would attain a $\pm 5\%$ precision for the measurement of stromal coherent transmittance (versus $\pm 9\%$ and $\pm 18\%$ for single measurements in 'Line' and 'Cross' mode, respectively). This precision could even be further improved to $\pm 3\%$ if analyzing 5 or more images in 'Line' or 'Cross' mode (see Figure 3).

Using a dataset of 83 normal corneas, we established normative measures of corneal transparency — based on the scattering mean-free path, $l_s$ (median = 570 µm, IDR: 270–2400 µm), in homogeneous stromal ROIs ($B_r$, IDR: 1.9–5.1), notably the fraction of transmitted coherent light, which considers stromal thickness, $T_{coh(stroma)}$ (median = 51%, IDR: 22–83%) — that may serve as further reference.

These values for transmitted coherent light, reflecting stromal transparency ($T_{coh(stroma)}$), are slightly lower than those for full corneal thickness previously reported in the literature at 840 nm or the closest available wavelength (see Table 1). This may be partly explained by our method being exclusively sensitive to the attenuation of coherent light. Note that our measurements are obtained with OCT, whose contrast arises from singly backscattered, or ballistic, photons in a few-µm-thick coherence plane; in other words, the detection (solid angle) is close to 0 steradian ($sr$). Previously reported values, on the other hand, were typically obtained with methods using a detector that is subtended by a larger, minimal solid angle, even for direct measurements. Indeed, the mean fraction of transmitted coherent light that we obtained for our group of normal corneas is compatible with literature values that were obtained with measurement methods whose detection (solid angle) was reduced to the vicinity of the ballistic direction (such as in [17, 18]). Another reason is that 41 of the 83 normal corneal images that were analyzed had saturation artifacts wider than 300 µm, which was associated with lower $T_{coh(stroma)}$ values (Figure 6). If eliminating these 41 images from our normative database, in other words, if only including the results from the 42 images with no and/or saturation artifacts narrower than 300 µm, the median fraction of transmitted coherent light obtained is raised to 70% (IDR: 34–87%).

*Table 1. Comparison with previously reported measurements of corneal transmittance, at 840 nm or the closest available wavelength [17–22]. Standard deviations for averaged transmittance values are given when available.*

| Year | Reference | Set-up | Wavelength (nm) | Solid angle (sr) | Measurement (sample size) | Corneal transmittance | Age (yr) Mean [range] | Age trend |
|------|-----------|--------|-----------------|------------------|---------------------------|-----------------------|------------------------|-----------|
| 1962 | Boettner & Wolter [19] | *in vitro* | 840 | $10^1$ (total)[*] | Average (6 in 9) | 95% | 35 [0.3–75] | No |
|      |           |        | 840 | $10^{-4}$ (direct)[†] | Single (best observed) | 80% | 4.5 | — |
|      |           |        |     | $10^{-4}$ (direct)[†] | Single *(near the av. of 8)* | 67% | 53 | — (*No*) |
| 1984 | Lerman [20] | *in vitro* | 750 | — | Single | 90% | 8 | — |
|      |           |        |     | — | Single | 77% | 24 |   |
|      |           |        |     | — | Single | 68% | 80 |   |
| 1990 | Beems & van Best [21] | *in vitro* | 700 | —[‡] | Average (8) | 96% SD ≤ 9% | 61 [22–87] | No |
| 1994 | van den Berg & Tan [22] | **in vivo** | 700 | $10^1$ (total)[*] | Average (10) | 94% | 51 [14–75] |   |
|      |           |        | 700 | $10^{-4}$ (direct)[†] | Average (10) | 89% | 51 [14–75] | No |
| 2010 | Peyrot *et al.* [17] | *in vitro* | 840 | $10^{-7}$ (direct) | Single *deswollen cornea* | 67% | — | — |
| 2013 | Crotti *et al.* [18] | *in vitro* | 840 | $10^{-7}$ (direct) | Single *mildly edematous cornea* | 48% | — | — |
| 2021 | Present study | **in vivo** | 840 | ~ 0 | Average (83) | 51%[§] IDR: [22–83]% | 31 [22–50] | No |
|      |           |        |     |   | Average (42)[‖] | 70%[§] IDR: [34–87]% | 32 [24–50] | No |

[*] Solid angle measured for an integrating sphere with an acceptance angle close to 180° (Ω = 2π[1-cos(170/2)] = 5.7 sr ≈ $10^1$ sr).
[†] Solid angle measured for an acceptance angle of 1° (Ω = π·tan(1/2)² ≈ $10^{-4}$ sr).
[‡] The acceptance angle was not specified. They used a 2.7 mm wide square photodiode implanted in the anterior chamber (if the distance between the front of this photodiode and the centre of the cornea was 1.5 mm, the acceptance angle and the solid angle were around 80 degrees and $10^0$ sr, respectively).
[§] Median coherent transmittance of the corneal stroma (the stromal ROI is defined in the present article). IDR stands for interdecile range, i.e, the range between the 10th percentile and the 90th percentile.
[‖] The reduced sample of n=42 corneas corresponds to OCT images with no saturation artifact or a saturation artifact narrower than 300 µm.

We found no significant correlation between age and central pachymetry (central corneal thickness, CCT) in our group of patients (that is relatively young due to the inclusion criteria having been registration for refractive surgery; age range from 22 to 50 years), as was previously reported for subjects younger than 50 years [23]. We also found no significant correlation between age and stromal coherent transmittance ($T_{coh(stroma)}$) at 840 nm, which is in agreement with prior full thickness transmittance measures [19, 21, 22] (note that a decrease in corneal light transmission with age was reported in [20] but later contradicted in [22]). Similarly, no age-related differences or correlations were previously found for corneal birefringence (which, as corneal transparency, is related to stromal structure) [24]. Further work will include a larger group of patients (including with various pathologies) with broader age ranges, enabling us to investigate the impact of age on *in vivo* near-infrared stromal transparency in subjects older than 50 years.

We did, however, find an impact of the grouping factor at the subject level for central corneal thickness (CCT) and stromal transmittance measurements: it explains 98% [96–99%] of the total variance of CCT (as has been previously described in [25]) and 55% [30–73%] of the total variance of $T_{coh(stroma)}$ measurements. The residual variance in $T_{coh(stroma)}$ may be explained by intra-subject variations, depending on each eye's medical history, as well as by the uncertainty related to single measurements. We noticed no significant difference in CCT or stromal transmittance between sides, that is, the right cornea and left cornea of the same subject, as reported for CCT in [23] and for corneal birefringence in [24].

While subjective and qualitative image inspection (and/or grading scale) routinely used by ophthalmologists in clinical practice is capable of distinguishing heterogeneous (abnormal; $B_r \gg 1$) corneas from homogeneous corneas ($B_r \sim 1$), it may fail to differentiate between those homogeneous corneas with compromised transparency (highly scattering; $l_s <$ corneal thickness) and those being transparent ($l_s \gg$ corneal thickness, as can be distinguished and classified by our algorithm). This is illustrated in Figure 7 using four representative clinical cases, by associating raw clinical SD-OCT images with their respective (z-) attenuation profiles and objective parameters. The low transparency measurements obtained for some normal corneas ($T_{coh(stroma)}$ between 13% and 30%; e.g. Figure 7B) raise the question of how a decrease in transmitted coherent light impacts patients' vision from a psychophysical point of view, since those low-coherent-transmittance corneas are still considered 'clinically normal' and subjects all have a best-corrected visual acuity of 20/20 (0.0 logMAR) or better. By creating a normative database of our parameters for pathological corneas, we shall be able to define thresholds and ranges of these physically relevant parameters in relation to clinically relevant indications, to provide a user-friendly and quantitative tool to ophthalmologists.

Future work will hence focus on the application of our approach to corneas affected by specific pathologies, including but not limited to Fuchs dystrophy, along with the study of the progression of the disease and post-operative follow-up.

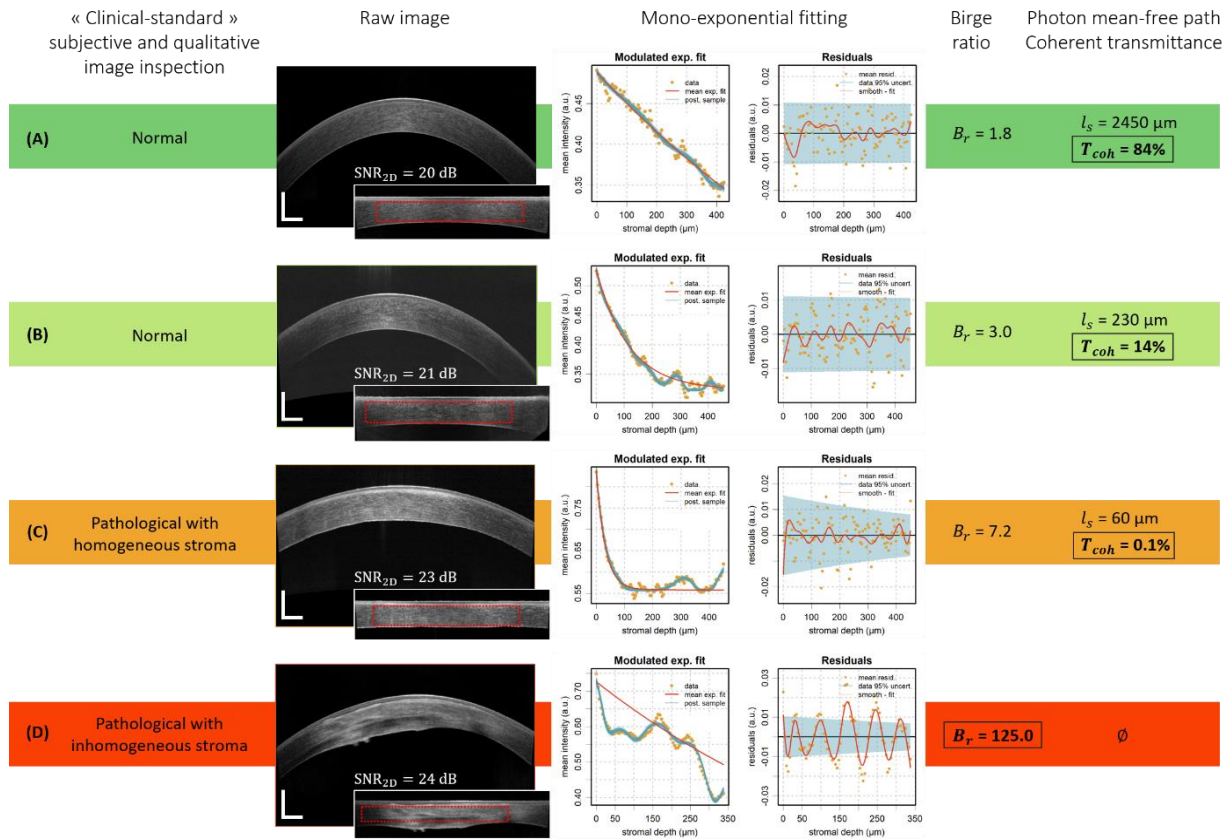

*Figure 7. Graphical representation of four representative clinical cases.* Each row comprises a SD-OCT cross-sectional 'Line' image of typical in vivo human cornea with corresponding flattened and normalized image (the apex-centered ROI appears as a red dotted frame), the associated mean intensity depth profile and mono-exponential fitting analysis. (A, B) Normal corneas from our n=83 sample with comparable Birge ratios, but (B) shows a faster decaying stromal depth profile and thus has a lower photon mean-free path ($l_s$) and associated fraction of transmitted coherent light ($T_{coh(stroma)}$). (C, D) Pathological corneas with compromised transparency as per "gold-standard" subjective and qualitative image inspection, with (C, Fuchs dystrophy) showing homogeneous scattering in the stroma ($B_r < 10$), resulting in a very fast in-depth decay and thus extremely low $T_{coh(stroma)}$, and (D, keratoconus) with visible heterogeneities (very heterogeneous scattering) in the stroma ($B_r \gg 10$); transparency assessment via a mono-exponential fitting of the in-depth profile is consequently inadequate.

## Acknowledgments

We gratefully acknowledge the contributions of Bathilde Rivière, Clara Faveau, and Astrid Minier who assisted with preliminary SD-OCT image analysis, as well as Hugo Lama and Faustine Dumont who helped with clinical data acquisition and export. We also thank Charline Pinna for the useful discussion related to the PCA analysis. The research was supported in part by the European Research Council (ERC) under the European Union's Horizon 2020 research and innovation program (Synergy grant agreement no. 610110, and Marie Skłodowska-Curie grant agreement no. 709104 (KI)), and by the LabEx PALM (ANR-10-LABX-0039-PALM; KP), as well as the "Banque Française des Yeux" (Prix BFY 2019; KP).